\documentclass[onecolumn,showpacs,showkeys,preprintnumbers,amsmath,amssymb,groupaddress]{revtex4}
\usepackage{mathbbold}
\usepackage{graphicx}% Include figure files
\usepackage{dcolumn}% Align table columns on decimal point
\usepackage{bm}% bold math

\begin{document}

\title{Quantum controlled phase gate based on two nonresonant quantum dots
trapped in two coupled photonic crystal cavities}
\author{Jian-Qi Zhang}
\author{Ya-Fei Yu}
\author{Xun-Li Feng}
\author{Zhi-Ming Zhang}
\email[Corresponding author Email: ]{zmzhang@scnu.edu.cn}
\date{\today}

\begin{abstract}
We propose a scheme for realizing two-qubit quantum phase gates with two
nonidentical quantum dots trapped in two coupled photonic crystal cavities
and driven by classical laser fields. During the gate operation, neither the
cavity modes nor the quantum dots are excited. The system can acquire
different phases conditional upon the different states of the quantum dots,
which can be used to realize the controlled phase gate.
\end{abstract}

\date{\today }
\maketitle
\preprint{APS/123-QED}
\affiliation{Key Laboratory of Photonic Information Technology of Guangdong Higher
Education Institutes, SIPSE $\&$ LQIT, South China Normal University,
Guangzhou 510006, China\\
}


\section{Introduction}

In recent years, there are great advancements on constructing the basic
components of quantum information processing (QIP) devices both in
experiments and theories \cite{01}. As the cavity quantum electrodynamics
(CQED) can manipulate the qubits efficiently, it has been become one of the
most promising approaches to realize the QIP devices \cite{02,03}. Although
the qubits in CQED can be atoms \cite{03}, ions \cite{05,06}, or quantum
dots (QDs) \cite{07}, the demonstrations of such basic building blocks of
the quantum on-chip network have relied on the atomic systems \cite{08,09,10}%
. Furthermore, a solid state implementation of these pioneering approaches
would open new opportunities for scaling the network into practical and
useful QIP systems \cite{01}. Among the proposed schemes based on solid
quantum devices, the systems of self-assembled QDs embedded in photonic
crystal (PC) nanocavities have been a kind of very promising systems. That
is not just because the strong QD-cavity interaction can be realized in
these systems \cite{11,12,13}, but also because both QDs and PC cavities are
suitable for monolithic on-chip integration.

However, there are two main challenges in this kind of systems. One is that
the variation in emission frequencies of the self-assembled QDs is large 
\cite{14}, the other is that the interaction between the QDs is difficult to
control \cite{15}. So far, there are several methods which have been used to
bring the emission frequencies of nonidentical QDs into the same, such as,
by using Stark shift tuning \cite{16} and voltage tuning \cite{17}. There
are also several solutions which have been used to control the interaction
between QDs, for instance, coherent manipulating coupled QDs \cite{15}, and
controlling the coupled QDs by Kondo effect \cite{18}. In experiments, the
tuning of individual QD frequencies has been achieved for two closely spaced
QDs in a PC cavity \cite{17}. However, there are few schemes about how to
achieve the controlled interaction and the controlled gate with the QDs
trapped in two coupled cavities.

Recently, Zheng proposed a scheme for implementing quantum gates by using
two atoms trapped in distant cavities connected by an optical fiber \cite%
{241}. But his proposal is based on two identical atoms, and there is no
directly coupling between the cavities. Motivated by this work, we propose a
scheme for realizing the controlled phase gate with two different QDs
trapped in two directly coupled PC cavities. The advantages of our scheme
are as follows. Firstly, it could be controlled by the external light
fields. Secondly, it can be realized in the case of large variation in
emission frequencies of the QDs. Thirdly, there is no cavity photon
population involved and the QDs are always in their ground states. Moreover,
our scheme does not require the condition that the coupling between QD and
cavity is smaller than that between cavities.

The organization of this paper is as follows. In Sec. \ref{sec2}, we
introduce the theoretical model and Hamiltonian. In Sec. \ref{sec3}, we
present the derivation of effective Hamiltonian. In Sec. \ref{sec4}, we show
the controlled phase gate. The discussion and conclusion is given in Sec. %
\ref{sec5}.

\section{Theoretical Model and Hamiltonian}

\label{sec2} We consider that two charged GaAs/AlGaAs QDs are placed in two
coupled single-mode PC cavities, which have the same frequency. Each dot has
two lower states $|g\rangle =|\uparrow \rangle $, $|f\rangle =|\downarrow
\rangle $ and a higher state $|e\rangle =|\uparrow \downarrow \Uparrow
\rangle $, here ($|\uparrow \rangle $, $|\downarrow \rangle $) and ($%
|\Uparrow \rangle $, $|\Downarrow \rangle $) denote the spin up and spin
down for electron and hole, respectively. The transitions $|g\rangle
\leftrightarrow |e\rangle $\ and $|f\rangle \leftrightarrow |e\rangle $ are
coupled to the vertical polarization and horizontal polarization lights,
respectively. Choosing the fields with the vertical polarization, the state $%
|f\rangle $ is not affected during the interactions, and only the transition 
$|g\rangle $ $\leftrightarrow $ $|e\rangle $ is coupled to the cavity mode
and classical laser field \cite{35}. Then the Hamiltonian for this model can
be written as: 
\begin{equation}
\begin{array}{rcl}
\hat{H} & = & \sum\limits_{j=A,B}(g_{j}a_{j}e^{i\Delta _{j}^{C}t}+\Omega
_{j}e^{i\Delta _{j}t})\sigma _{j}^{+}+\nu a_{A}^{+}a_{B}+H.c.%
\end{array}
\label{eq01-1}
\end{equation}%
where$\ \sigma _{j}^{+}=|e\rangle _{j}\langle g|$, $g_{j}$ is the coupling
constant between the cavity $j$ and QD $j$, $\Omega _{j}$ are the Rabi
frequencies of the laser fields, the detunings are $\Delta _{j}^{C}$, and $%
\Delta _{j}$, respectively, $a_{j}^{\dag }$ and $a_{j}$ are the creation and
annihilation operators for the $j$th cavity mode, $\nu $ is the coupling
strength between the two cavity modes (see FIG.1 ).

\section{Derivation of effective Hamiltonian}

\label{sec3} Introducing new annihilation operators $c_{1}$ and $c_{2}$, and
defining $a_{A}=\frac{1}{\sqrt{2}}(c_{1}+c_{2})$, $a_{B}=\frac{1}{\sqrt{2}}%
(c_{2}-c_{1})$, and $\Delta _{j}^{C}=\Delta _{j}+\delta $, the Hamiltonian
can be rewritten as 
\begin{equation}
\begin{array}{rcl}
\hat{H}_{int} & = & \hat{H}_{0}+\hat{H}_{i} \\ 
\hat{H}_{0} & = & \nu (c_{2}^{+}c_{2}-c_{1}^{+}c_{1}), \\ 
\hat{H}_{i} & = & \sum\limits_{j=A,B}[\frac{1}{2}g_{j}(c_{1}+c_{2})e^{i(%
\Delta _{j}+\delta )t}+\Omega _{j}e^{i\Delta _{j}t}]\sigma _{j}^{+} \\ 
& + & H.c.%
\end{array}
\label{eq01-2}
\end{equation}

With the application of the unitary transformation $e^{iH_{0}t}$, the
Hamiltonian takes the form:

\begin{equation}
\begin{array}{rcl}
\hat{H}_{I} & = & [\frac{1}{2}g_{A}(c_{2}e^{i(\Delta _{A}+\delta -\nu
)t}+c_{1}e^{i(\Delta _{A}+\delta +\nu )t})+\Omega _{A}e^{i\Delta
_{A}t}]\sigma _{A}^{+} \\ 
& + & [\frac{1}{2}g_{B}(c_{2}e^{i(\Delta _{B}+\delta -\nu
)t}-c_{1}e^{i(\Delta _{B}+\delta +\nu )t})+\Omega _{B}e^{i\Delta
_{B}t}]\sigma _{B}^{+} \\ 
& + & H.c.%
\end{array}
\label{eq01-3}
\end{equation}

\begin{figure}[tbph]
\includegraphics[width=8cm]{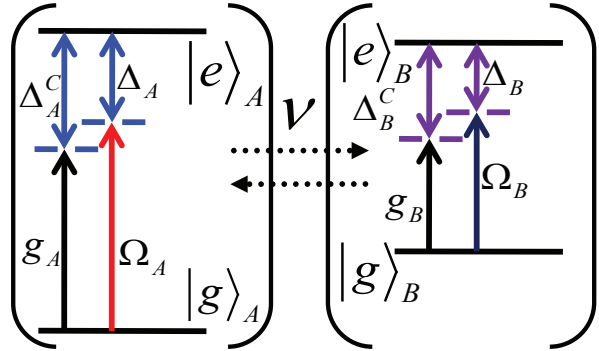}
\caption{Schematic diagram of the system. Each dot is trapped in its
corresponding cavity and driven by a light field. Photons can hop between
the cavities.}
\label{fig 1}
\end{figure}

Now, we will use the method proposed in Ref. \cite{24,28} to derive the
effective Hamiltonian for this system. With $|\Delta _{j}|,|\Delta
_{j}^{^{\prime }}|\gg \nu ,\delta ,|g_{j}|,|\Omega _{j}|$ assumed, the
probability for QDs absorbing photons from the light field and being excited
will be ignored, and the excited state of QD can be adiabatically
eliminated. Thus we can obtain the effective Hamiltonian:

\begin{equation}
\begin{array}{rcl}
\hat{H}_{eff-1} & = & -c_{2}(\lambda _{A,2}\sigma _{A}^{-}\sigma
_{A}^{+}-\lambda _{B,2}\sigma _{B}^{-}\sigma _{B}^{+})e^{i(\delta -\nu )t}
\\ 
& - & c_{1}(\lambda _{A,1}\sigma _{A}^{-}\sigma _{A}^{+}+\lambda
_{B,1}\sigma _{B}^{-}\sigma _{B}^{+})e^{i(\delta +\nu )t} \\ 
& - & (k_{A}\sigma _{A}^{-}\sigma _{A}^{+}-k_{B}\sigma _{B}^{-}\sigma
_{B}^{+})c_{1}c_{2}^{+}\sigma _{A}^{-}\sigma _{A}^{+}e^{-i2vt} \\ 
& + & H.c. \\ 
& - & (l_{A,1}c_{1}^{+}c_{1}+l_{A,2}c_{2}^{+}c_{2}+l_{A,3})\sigma
_{A}^{-}\sigma _{A}^{+} \\ 
& - & (l_{B,1}c_{1}^{+}c_{1}+l_{B,2}c_{2}^{+}c_{2}+l_{B,3})\sigma
_{B}^{-}\sigma _{B}^{+}.%
\end{array}
\label{eq03}
\end{equation}%
where%
\[
\begin{array}{l}
\lambda _{j,1}=\dfrac{g_{j}\Omega _{j}^{\ast }}{4}(\frac{1}{\Delta
_{j}+\delta +\nu }+\frac{1}{\Delta _{j}}); \\ 
\lambda _{j,2}=\dfrac{g_{j}\Omega _{j}^{\ast }}{4}(\frac{1}{\Delta
_{j}+\delta -\nu }+\frac{1}{\Delta _{j}}); \\ 
k_{j}=\dfrac{|g_{j}|^{2}}{8}(\frac{1}{\Delta _{j}+\delta +\nu }+\frac{1}{%
\Delta _{j}+\delta -\nu }); \\ 
l_{j,1}=\frac{|g_{j}|^{2}}{4(\Delta _{j}+\delta -\nu )};l_{j,2}=\frac{%
|g_{j}|^{2}}{4(\Delta _{j}+\delta +\nu )}; \\ 
l_{j,3}=\frac{|\Omega _{j}|^{2}}{\Delta _{j}};\Delta _{1}=\delta -\nu
;\Delta _{2}=\delta +\nu .%
\end{array}%
\]

Under the condition $\delta +\nu ,\delta -\nu ,2v\gg \lambda _{j,1},\lambda
_{j,2},k_{j}$, the new bosonic modes cannot exchange energy with each other
and with the classical fields, the coupling between the two cavities can be
much larger than the one between QD and cavity. Moreover, the couplings
between the bosonic modes and the classical fields lead to energy shifts
which are only depending upon the number of QDs in the state $|g\rangle $,
while the couplings between different bosonic modes cause energy shifts
depending upon both the excitation numbers of the modes and the number of
QDs in the state $|g\rangle $. Then the effective Hamiltonian takes the form:

\begin{equation}
\begin{array}{rcl}
\hat{H}_{eff-2} & = & \frac{1}{\delta -\nu }(\lambda _{A,2}\sigma
_{A}^{-}\sigma _{A}^{+}-\lambda _{B,2}\sigma _{B}^{-}\sigma _{B}^{+}) \\ 
& \ast & (\lambda _{A,2}^{\ast }\sigma _{A}^{-}\sigma _{A}^{+}-\lambda
_{B,2}^{\ast }\sigma _{B}^{-}\sigma _{B}^{+}) \\ 
& + & \frac{1}{\delta +\nu }(\lambda _{A,1}\sigma _{A}^{-}\sigma
_{A}^{+}+\lambda _{B,1}\sigma _{B}^{-}\sigma _{B}^{+}) \\ 
& \ast & (\lambda _{A,1}^{\ast }\sigma _{A}^{-}\sigma _{A}^{+}+\lambda
_{B,1}^{\ast }\sigma _{B}^{-}\sigma _{B}^{+}) \\ 
& + & \frac{1}{2v}(k_{A}\sigma _{A}^{-}\sigma _{A}^{+}-k_{B}\sigma
_{B}^{-}\sigma _{B}^{+})^{2}(c_{1}^{+}c_{1}-c_{2}^{+}c_{2}) \\ 
& - & (l_{A,1}c_{1}^{+}c_{1}+l_{A,2}c_{2}^{+}c_{2}+l_{A,3})\sigma
_{A}^{-}\sigma _{A}^{+} \\ 
& - & (l_{B,1}c_{1}^{+}c_{1}+l_{B,2}c_{2}^{+}c_{2}+l_{B,3})\sigma
_{B}^{-}\sigma _{B}^{+}.%
\end{array}%
\end{equation}%
It shows, during the interaction, the excitation numbers of the bosonic
modes $c_{1}$ and $c_{2}$ are conserved, so does the one for the cavity
modes. Assume that the initial state for two cavity modes is in the vacuum
state, the new bosonic modes will be in the vacuum state during the
evolution. In this situation, the effective Hamiltonian reduces to

\begin{equation}
\begin{array}{rcl}
\hat{H}_{eff} & = & \frac{1}{\delta -\nu }(\lambda _{A,2}\sigma
_{A}^{-}\sigma _{A}^{+}-\lambda _{B,2}\sigma _{B}^{-}\sigma _{B}^{+}) \\ 
& \ast & (\lambda _{A,2}^{\ast }\sigma _{A}^{-}\sigma _{A}^{+}-\lambda
_{B,2}^{\ast }\sigma _{B}^{-}\sigma _{B}^{+}) \\ 
& + & \frac{1}{\delta +\nu }(\lambda _{A,1}\sigma _{A}^{-}\sigma
_{A}^{+}+\lambda _{B,1}\sigma _{B}^{-}\sigma _{B}^{+}) \\ 
& \ast & (\lambda _{A,1}^{\ast }\sigma _{A}^{-}\sigma _{A}^{+}+\lambda
_{B,1}^{\ast }\sigma _{B}^{-}\sigma _{B}^{+}) \\ 
& - & l_{A,3}\sigma _{A}^{-}\sigma _{A}^{+}-l_{B,3}\sigma _{B}^{-}\sigma
_{B}^{+}.%
\end{array}
\label{eq302}
\end{equation}%
This equation can be understood as follows. With the laser field acting, QDs
will take place the Stark shifts and acquire the virtual excitation, and the
virtual excitation will induce the coupling between the vacuum bosonic modes
and classical fields. As the Stark shifts are nonlinear in the number of the
QDs in the state $|g\rangle $, the system can acquire a phase conditional
upon the number of the QDs in the state $|g\rangle $.

\section{The Controlled Phase gate}

\label{sec4} Now, we will show how to construct the controlled phase gate in
this system. First of all, the inforamtion of the system is encoded in the
states $|g\rangle $ and $|f\rangle $. Then, according to the effective
Hamiltonian (\ref{eq302}), the evolution for states \{$|ff\rangle $, $%
|fg\rangle $, $|gf\rangle $, and $|gg\rangle $\} can be written as: 
\begin{equation}
\left\{ 
\begin{array}{rcl}
|ff\rangle & \rightarrow & |ff\rangle , \\ 
|fg\rangle & \rightarrow & e^{-i\Phi _{A}t}|fg\rangle , \\ 
|gf\rangle & \rightarrow & e^{-i\Phi _{B}t}|gf\rangle , \\ 
|gg\rangle & \rightarrow & e^{-i(\Phi _{A}+\Phi _{B}+\eta )t}|gg\rangle .%
\end{array}%
\right.  \label{eq109}
\end{equation}%
where%
\begin{equation}
\left\{ 
\begin{array}{rcl}
\Phi _{A} & = & \frac{|\lambda _{A,2}|^{2}}{\delta -\nu }+\frac{|\lambda
_{A,1}|^{2}}{\delta +\nu }-l_{A,3}, \\ 
\Phi _{B} & = & \frac{|\lambda _{B,2}|^{2}}{\delta -\nu }+\frac{|\lambda
_{B,1}|^{2}}{\delta +\nu }-l_{B,3}, \\ 
\eta & = & \frac{1}{\delta +\nu }(\lambda _{A,1}\lambda _{B,1}^{\ast
}+\lambda _{A,1}^{\ast }\lambda _{B,1}) \\ 
& - & \frac{1}{\delta -\nu }(\lambda _{A,2}\lambda _{B,2}^{\ast }+\lambda
_{A,2}^{\ast }\lambda _{B,2}). \\ 
& = & 2(\frac{|\lambda _{A,1}\lambda _{B,1}|\cos \theta _{1}}{\delta +\nu }+%
\frac{|\lambda _{A,2}\lambda _{B,2}|\cos \theta _{2}}{v-\delta }),%
\end{array}%
\right.
\end{equation}%
$\theta _{1}$ and $\theta _{2}$ are the arguments of $\lambda _{A,1}\lambda
_{B,1}^{\ast }$ and $\lambda _{A,2}\lambda _{B,2}^{\ast }$, respectively.

With the application of single-qubit operations\cite{29} 
\begin{equation}
\left\{ 
\begin{array}{rcl}
|g_{A}\rangle & \rightarrow & e^{i\Phi _{A}t}|g_{A}\rangle , \\ 
|g_{B}\rangle & \rightarrow & e^{i\Phi _{B}t}|g_{B}\rangle ,%
\end{array}%
\right.  \label{eq110}
\end{equation}%
Eq.(\ref{eq109}) will transform into%
\begin{equation}
\left\{ 
\begin{array}{rcl}
|ff\rangle & \rightarrow & |ff\rangle , \\ 
|fg\rangle & \rightarrow & |fg\rangle , \\ 
|gf\rangle & \rightarrow & |gf\rangle , \\ 
|gg\rangle & \rightarrow & e^{-i\eta t}|gg\rangle .%
\end{array}%
\right.  \label{eq111}
\end{equation}%
It is clearly, with the choice of $\eta t=\pi $, this transformation
corresponds to the quantum controlled phase $\pi $ gate operation, in which
if and only if both controlling and controlled qubits are in the states $%
|g\rangle $, there will be an additional phase $\pi $ in the system.

\section{Discussion and Conclusion}

\label{sec5} In order to confirm the validity of the proposal, we takes
controlled phase $\pi $ gate (C-Z gate) as an example to disscuss the
realizability in the experiment. According to experimentally achievable
parameters in the system of QDs embedded in a single-mode cavity \cite{35,36}%
, the coupling constant between cavity and QD is $g\sim 0.1meV$, the decay time for cavity is $%
\tau _{c}\sim 1ns$, and the energy relaxation time of the excited state is $%
\tau _{e}\sim 1.4ns$. With the choices of the coupling constants and
detunings as follows: $g_{A}=0.1meV $, $g_{B}=0.08meV$, $\Omega _{A}=10meV$, 
$\Omega _{B}=13.75meV$, $\Delta _{A}=200meV$, $\Delta _{B}=220meV$, $\delta
=2g_{A}$, $\nu =12g_{A}$, we have $\lambda _{j,1}\simeq \lambda _{j,2}\sim
0.0025meV$, which satisfy the approximation conditions mentioned above. The
calculations show that i) the max-occupation probability of the excited
state is $\max (P_{e})\sim 1/256(\simeq \max [\Omega _{j}^{2}/\Delta
_{j}^{2}])$, and thus, the effective energy relaxation time is $t_{e}\sim
358ns(\simeq \tau _{e}/P_{e})$. ii) the occupation probability of the photon
is $P_{c}\sim 1/900(\simeq \max [\lambda _{j,1}^{2},\lambda
_{j,2}^{2}]/\delta ^{2})$, so the effective decay time is $t_{c}\sim
900ns(\simeq \tau _{c}/P_{c})$, iii) the required effective interaction time
for the C-Z gate is $t\sim 50ns(\simeq \pi /(2\epsilon ))$. Therefore, it is
possible to perform several C-Z gates within the decoherence time $\min
[t_{e},t_{c}]\sim 358ns$.

In summary, we have shown a protocol that two nonidentical QDs trapped in
two coupled PC cavities can be used to construct the two-qubit controlled
phase gate with the application of the classical light fields. During the
gate operation, none of the QDs is in the excited state, and both of the
cavities are in the vacuum state. The distinct advantages of the proposed
scheme are as follows: firstly, it is controllable; secondly, during the
gate operation, there is no cavity photon population involved and the QDs
are always in their ground states; finally, as the QDs are non-identical and
the coupling between the two cavities can be much larger than the one
between QD and cavity, it is more practical. Therefore, we can use this
scheme to construct a kind of solid-state controllable quantum logical
devices. In addition, as the controlled phase gate is a universal gate, this
system can also realize the controlled entanglement and interaction between
the two nonidentical QDs trapped in two coupled cavities.

\section*{Acknowledgement}

This work was supported by the National Natural Science Foundation of China
(Grant No. 60978009) and the National Basic Research Program of China (Grant
Nos. 2009CB929604 and 2007CB925204).

\end{document}